\documentclass{aa}  
\usepackage{natbib}
\usepackage{CJK}

\usepackage{graphicx}
\usepackage{txfonts}
\usepackage{xcolor}
\usepackage{multirow}
\usepackage{hyperref}

\def\h2{\ion{H}{II}}

\begin{document}

   \title{Discovery of ammonia (9,6) masers in two high-mass star-forming regions}

  % \subtitle{}

   \author{{\protect\begin{CJK*}{UTF8}{gkai}Y. T. Yan (闫耀庭)\protect\end{CJK*}} \inst{\ref{inst.mpifr}}\fnmsep\thanks{Member of the International Max Planck Research School (IMPRS) for Astronomy and Astrophysics at the universities of Bonn and Cologne.}
         \and C. Henkel\inst{\ref{inst.mpifr},\ref{inst.KingAbdulazizU},\ref{inst.xao}}
         \and K. M. Menten\inst{\ref{inst.mpifr}}
         \and {\protect\begin{CJK*}{UTF8}{gkai}Y. Gong (龚\protect\end{CJK*}\protect\begin{CJK*}{UTF8}{bkai}龑\protect\end{CJK*})} \inst{\ref{inst.mpifr}}
         \and J. Ott\inst{\ref{inst.nrao}}
         \and T. L. Wilson\inst{\ref{inst.mpifr}}
         \and A. Wootten\inst{\ref{inst.nrao}}
         \and A. Brunthaler\inst{\ref{inst.mpifr}}
         \and {\protect\begin{CJK*}{UTF8}{gkai}J. S. Zhang (张江水)\protect\end{CJK*}}\inst{\ref{inst.gzhu}}
         \and {\protect\begin{CJK*}{UTF8}{gkai}J. L. Chen (陈家梁)\protect\end{CJK*}}\inst{\ref{inst.gzhu}}
         \and {\protect\begin{CJK*}{UTF8}{gkai}K. Yang (杨楷)\protect\end{CJK*}}\inst{\ref{inst.nju},\ref{inst.keylaboratorynj}}
         }

   \institute{
\label{inst.mpifr}Max-Planck-Institut f\"{u}r Radioastronomie, Auf dem H\"{u}gel 69, 53121 Bonn, Germany\\  \email{yyan@mpifr-bonn.mpg.de}
\and\label{inst.KingAbdulazizU}Astronomy Department, Faculty of Science, King Abdulaziz University, P.~O.~Box 80203, Jeddah 21589, Saudi Arabia
\and\label{inst.xao}Xinjiang Astronomical Observatory, Chinese Academy of Sciences, 830011 Urumqi, PR China
\and\label{inst.nrao}National Radio Astronomy Observatory, 520 Edgemont Road, Charlottesville, VA 22903-2475, USA
\and\label{inst.gzhu}Center for Astrophysics, Guangzhou University, 510006 Guangzhou, People's Republic of China
\and\label{inst.nju}School of Astronomy and Space Science, Nanjing University, 163 Xianlin Avenue, Nanjing 210023, People's Republic of China
\and\label{inst.keylaboratorynj}Key Laboratory of Modern Astronomy and Astrophysics (Nanjing University), Ministry of Education, Nanjing 210023, People's Republic of China
             }

   \date{Received 13 December 2021 / Accepted 30 December 2021}

% \abstract{}{}{}{}{} 
% 5 {} token are mandatory
 
  \abstract
  % context heading (optional)
  % {} leave it empty if necessary  
{Molecular maser lines are signposts of high-mass star formation, probing the excitation and kinematics of very compact regions in the close environment of young stellar objects and providing useful targets for trigonometric parallax measurements.}
  % aims heading (mandatory)
{Only a few NH$_{3}$ (9,6) masers are known so far, and their origin is still poorly understood. Here we aim to find new NH$_{3}$ (9,6) masers to provide a better observational basis for studying their role in high-mass star-forming regions.}
  % methods heading (mandatory)
{We carried out NH$_{3}$ (9,6) observations toward Cepheus A and G34.26$+$0.15 with the Effelsberg 100-meter telescope (beam size 49\arcsec) and the Karl G. Jansky Very Large Array (JVLA; beam size about 1$\farcs$2).}
  % results heading (mandatory)
{We discovered new NH$_{3}$ (9,6) masers in Cep A and G34.26$+$0.15, which increases the number of known high-mass star-forming regions hosting NH$_{3}$ (9,6) masers from five to seven. Long-term monitoring (20 months) at Effelsberg shows that the intensity of the (9,6) maser in G34.26$+$0.15 is decreasing, while the Cep A maser remains stable. Compared to the Effelsberg data and assuming linear variations between the epochs of observation, the JVLA data indicate no missing flux. This suggests that the NH$_3$ (9,6) emission arises from single compact emission regions that are not resolved by the interferometric measurements. As JVLA imaging shows, the NH$_{3}$ (9,6) emission in Cep A originates from a sub-arcsecond-sized region, slightly to the west ($0\farcs28 \pm 0\farcs10$)  of the peak position of the 1.36\,cm continuum object, HW2. In G34.26$+$0.15, three NH$_{3}$ (9,6) maser spots are observed: one is close to the head of the cometary ultracompact \h2 region C, and the other two are emitted from a compact region to the west of the hypercompact \h2 region A.}
  % conclusions heading (optional), leave it empty if necessary 
{The newly found (9,6) masers appear to be related to outflows. The higher angular resolution of JVLA and very long baseline interferometry observations are needed to provide more accurate positions and constraints for pumping scenarios.}

\keywords{Masers -- ISM: clouds -- ISM: individual objects: Cep A, G34.26$+$0.15 -- ISM: \h2 regions -- Radio lines: ISM}

\maketitle

\section{Introduction}
\label{introduction}

Since its discovery more than five decades ago \citep{1968PhRvL..21.1701C}, ammonia (NH$_3$) has been a most valuable molecule for investigating the physical properties of molecular clouds \citep[e.g.,][]{1983ARA&A..21..239H}. While  thermally excited transitions in the centimeter-wavelength inversion transitions of ammonia are regarded as a reliable thermometer  of molecular clouds \citep[e.g.,][]{1983A&A...122..164W,1988MNRAS.235..229D}, ammonia masers have attracted attention since the first detection of maser action in the $(J,K)$~=~(3,3) metastable ($J=K$) line toward the massive star-forming region W33 \citep{1982A&A...110L..20W}. Subsequent observations have led to the detection of new metastable ammonia masers, including $^{15}$NH$_3$ (3,3) \citep{1986A&A...160L..13M}, NH$_{3}$ (1,1) \citep{1996ApJ...457L..47G}, NH$_{3}$ (2,2) \citep{2018ApJ...869L..14M}, NH$_{3}$ (5,5) \citep{1992A&A...256..618C}, NH$_{3}$ (6,6) \citep{2007A&A...466..989B}, NH$_{3}$ (7,7), NH$_{3}$ (9,9), and NH$_{3}$ (12,12) \citep{2013A&A...549A..90H}. These have led to the discovery of metastable maser lines in 22 different regions \citep{1986A&A...160L..13M,1987A&A...173..352M,1988A&A...206L..26W,1990A&A...229L...1W,1991ApJ...373L..13P,1992A&A...256..618C,1993LNP...412..123W,1994ApJ...428L..33M,1995ApJ...439L...9K,1995ApJ...450L..63Z,1999ApJ...527L.117Z,2007MNRAS.382L..35W,2008ApJ...680.1271H,2009ApJ...706.1036G,2011ApJ...739L..16B,2011MNRAS.418.1689U,2011MNRAS.416.1764W,2012ApJ...745L..30W,2013A&A...549A..90H,2014ApJ...782...83H,2016ApJ...826..189M,2018ApJ...869L..14M,2019ApJ...887...79H,2020ApJ...898..157M,2021ApJ...923..263T}. Compared with the metastable ammonia masers, detected non-metastable ($J>K$) ammonia maser transitions are more numerous. The first highly excited non-metastable ammonia maser was detected by \citet{1986ApJ...300L..79M} in the $(J,K)$~=~(9,6) and (6,3) lines. Thereafter, many other NH$_3$ non-metastable inversion transition lines have been identified as masers, including the (5,3), (5,4), (6,1), (6,2), (6,4), (6,5), (7,3), (7,4), (7,5) (7,6), (8,3), (8,4), (8,5), (8,6), (9,3), (9,4), (9,5), (9,7), (9,8), (10,7), (10,8), (10,9), and (11,9) transitions \citep[e.g.,][]{1987A&A...173..352M,1988A&A...201..123M,2007MNRAS.382L..35W,2013A&A...549A..90H,2020ApJ...898..157M}. Except for the NH$_{3}$ (3,3) masers proposed to be associated with four supernova remnants \citep{2016ApJ...826..189M}, almost all the other ammonia masers are detected in high-mass star-forming regions (HMSFRs). However, while many HMSFRs host water (H$_2$O), hydroxyl (OH), or methanol (CH$_3$OH) masers, ammonia masers are quite rare in these sources, and the role that the environment of a young high-mass star plays in their excitation remains unclear. Therefore, dedicated searches for ammonia masers in HMSFRs are indispensable in regard to their overall incidence and association with different environments, which can provide additional constraints on the pumping mechanism of ammonia masers.

So far, a total of 32 NH$_3$ inversion transitions ($\Delta K$~=~0 and $\Delta J$~=~0) have been identified as masers. Among these, and despite arising from energy levels as high as 1090~K above the ground state, the NH$_{3}$ (9,6) maser stands out as being the strongest and most variable one in W51-IRS2 \citep[e.g.,][]{2013A&A...549A..90H}. Maser emission in this line has only been detected in five HMSFRs, W51, NGC7538, W49, DR21 (OH) \citep{1986ApJ...300L..79M}, and Sgr B2(N) \citep{2020ApJ...898..157M}. The NH$_{3}$ (3,3) masers are thought to be collisionally excited  \citep[e.g.,][]{1990MNRAS.244P...4F,1994ApJ...428L..33M}; in contrast, the pumping mechanism of NH$_{3}$ (9,6) masers is less well constrained \citep{1986ApJ...300L..79M}. \citet{1991ApJ...378..445B} have studied ortho-ammonia and found that it could possibly pump the (6,3) inversion line, but they did not extend their model to the (9,6) transition due to the fact that collision rates are only known for inversion levels up to $J~=~6$ (e.g., \citealt{1988MNRAS.235..229D}). 

NH$_3$ (9,6) masers are found to be strongly variable, similar to H$_2$O masers (\citealt{1986ApJ...300L..79M,1991ApJ...373L..13P,2013A&A...549A..90H}). In W51-IRS2, \citet{2013A&A...549A..90H} found that the (9,6) line showed significant variation in line shape within a time interval of only two days. Mapping of the (9,6) maser toward W51 with very long baseline interferometry (VLBI) suggests that the masers are closer to the H$_2$O masers than to the OH masers or to ultracompact (UC) \h2 regions \citep{1991ApJ...373L..13P}. While  \citet{2013A&A...549A..90H} and \citet{2015A&A...573A.109G} showed that the SiO and NH$_3$ masers in W51-IRS2 are very close to each other, their positions, differing by 0$\farcs$065~($\sim$0.015 pc), do not fully coincide. 

In this paper we report the discovery of NH$_3$ (9,6) masers in two HMSFRs, Cepheus A and G34.26$+$0.15. This increases the number of (9,6) maser detections in our Galaxy from five to seven. In Sect. \ref{observations} observations with the Effelsberg 100-meter telescope and the Karl G. Jansky Very Large Array (JVLA) are described. Results are presented in Sect. \ref{results}. The morphology of Cep A and G34.26$+$0.15 as well as a comparison of the emission distributions of different tracers with the NH$_3$ (9,6) masers are presented in Sect. \ref{disscussion}. Our main results are summarized in Sect. \ref{summary}.

\section{Observations and data reduction}
\label{observations}

\subsection{Effelsberg observations and data reduction}
\label{EffelsbergObservations}

The NH$_3$ (9,6) line was observed toward Cep A and G34.26$+$0.15 with the 100-meter Effelsberg telescope\footnote{Based on observations with the 100-meter telescope of the MPIfR (Max-Planck-Institut f\"{u}r Radioastronomie) at Effelsberg.} in 2020 January and 2021 February, July, and August. The S14mm double beam secondary focus receiver was employed. The full width at half maximum (FWHM) beam size is 49$\arcsec$ at 18.5 GHz, the frequency of the target line. The observations were performed in position switching mode, and the off position was 10$\arcmin$ in azimuth away from the source. For observations made before 2021 August, we used a spectrometer that covered 2\,GHz wide backends with a channel width of 38.1 kHz, corresponding to $\sim$0.62~km~s$^{-1}$ at the line's rest frequency, 18.49939 GHz \citep{1975ApJS...29...87P}. A high spectral resolution backend with 65536 channels and a bandwidth of 300 MHz was employed in 2021 August, providing a channel width of 0.07~km s$^{-1}$ at 18.5 GHz. Pointing was checked every 2 hours using 3C~286 or NGC~7027. Focus calibrations were done at the beginning of the observations and during sunset and sunrise toward the abovementioned pointing sources. The system temperatures were 100--130 K on a main-beam brightness temperature, $T_{\rm MB}$, scale. This flux density was calibrated assuming a $T_{\rm MB}/S$ ratio of 1.95 K/Jy, derived from continuum cross scans of NGC 7027 (the flux density was adopted from \citealt{1994A&A...284..331O}). Calibration uncertainties are estimated to be $\sim10$\%. 

We used the GILDAS/CLASS\footnote{https://www.iram.fr/IRAMFR/GILDAS/} package \citep{2005sf2a.conf..721P} to reduce the spectral line data. A first-order polynomial was subtracted from each spectrum for baseline removal. 

\subsection{JVLA observations and data reduction}
\label{JVLAObservations}

Observations of the NH$_3$ (9,6) line toward Cep A and G34.26$+$0.15 were obtained on 2021 July 13 with the JVLA of the National Radio Astronomy Observatory\footnote{The National Radio Astronomy Observatory is a facility of the National Science Foundation operated under cooperative agreement by Associated Universities, Inc.} (NRAO) in the C configuration (project ID:  21A-157, PI: Yaoting Yan). We employed 27 antennas for the observations. The primary beam of the JVLA antennas is $150^{\prime \prime}$ (FWHM) at 18.5 GHz. A mixture of mixed three-bit and eight-bit samplers were used to perform the observations. For the NH$_3$ (9,6) line observations, we used one subband with the eight-bit sampler covering a bandwidth of 16 MHz with full polarization, eight recirculations, and four baseline board pairs (BIBPs) to provide a velocity range of 260~km~s$^{-1}$ with a channel spacing of ~0.13~km~s$^{-1}$. Two additional subbands of bandwidth 16 MHz were used to cover the NH$_3$ (8,5) and (10,7) lines. The three-bit sampler with 32 subbands, each with a bandwidth of 128 MHz to cover a total range of 4 GHz between 20--24 GHz, was used to measure the continuum emission. 3C~286 with a flux density of 2.89 Jy at 18.5 GHz \citep{2013ApJS..204...19P} was used as a calibrator for pointing, flux density, bandpass, and polarization. J2230+6946 and J1851+0035 served as gain calibrators for Cep A and G34.26$+$0.15, respectively. The on-source times were 4$^m$30$^s$ and 4$^m$50$^s$ toward Cep A and G34.26$+$0.15, respectively. 

Data from two antennas were lost due to technical issues. The data from the remaining 25 antennas were reduced through the Common Astronomy Software Applications package (CASA\footnote{https://casa.nrao.edu/}; \citealt{2007ASPC..376..127M}). We calibrated the data with the JVLA CASA calibration pipeline using CASA 6.1.2. The results were obtained after flagging data that contain artifacts. We inspected the phase, amplitude, and bandpass variations of the calibrated visibility data to search for additional artifacts before imaging. Then, the \textit{uvcontsub} task in CASA was used to separate the calibrated visibilities into two parts, one with line-only data and the other with the continuum data. The \textit{tclean} task with a cell size of 0$\farcs$2 and Briggs weighting with robust=0 was used to produce the images of spectral line and continuum emission. The synthesized beams for NH$_3$ (9,6) are $1\farcs47 \times 0\farcs99$ at P.A.~=~$58\fdg79$ and $1\farcs33 \times 1\farcs06$ at P.A.~=~$5\fdg36$ toward Cep A and G34.26$+$0.15, respectively. For the 1.36\,cm (20--24 GHz) continuum emission, the synthesized beams are $1\farcs08 \times 0\farcs67$ at P.A.~=~$60\fdg64$ and $0\farcs95 \times 0\farcs71$ at P.A.~=~$5\fdg91$ toward Cep A and G34.26$+$0.15. The typical absolute astrometric accuracy of the JVLA is $\sim$10\% of the synthesized beam\footnote{https://science.nrao.edu/facilities/vla/docs/manuals/oss/performance-/positional-accuracy}. The flux density scale calibration accuracy is estimated to be within 15\%.

\begin{figure}[h]
\center
        \includegraphics[height=290pt]{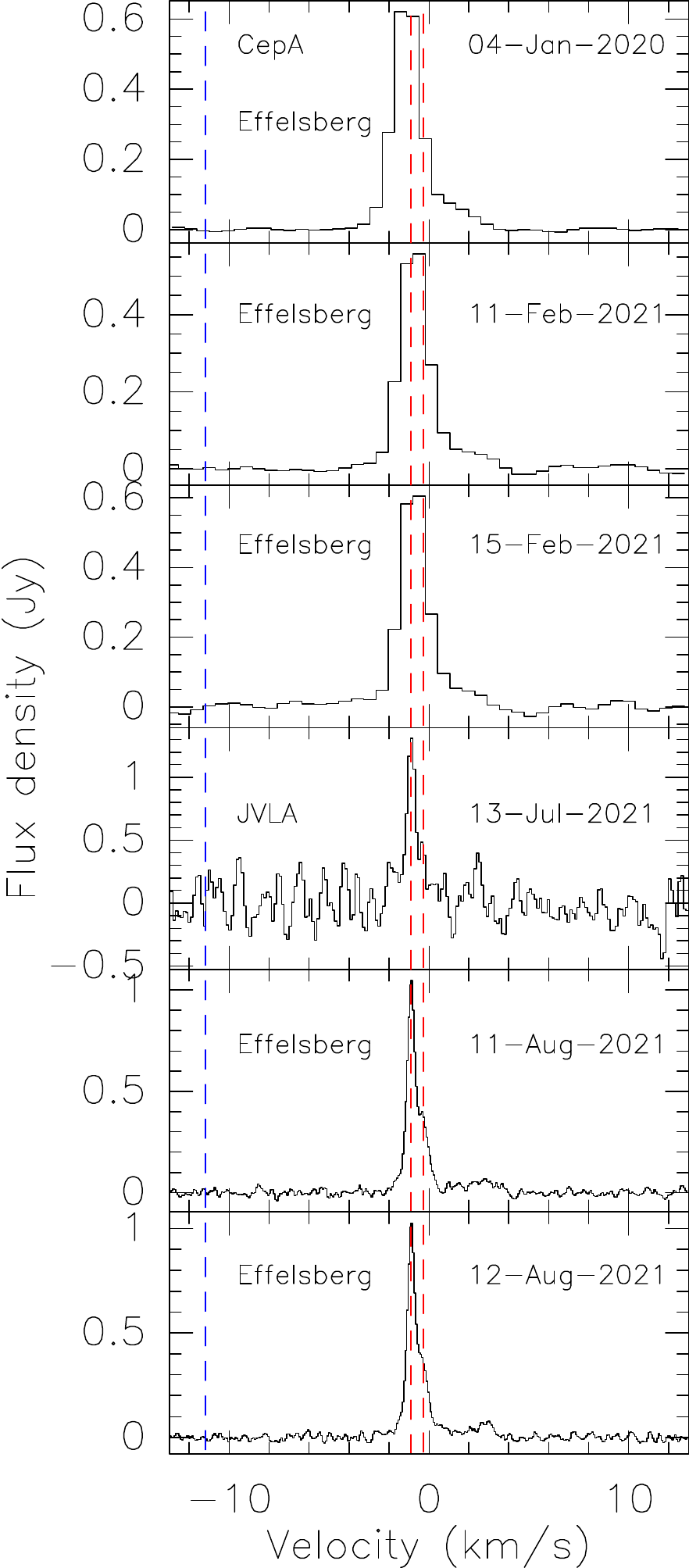}
        \includegraphics[height=290pt]{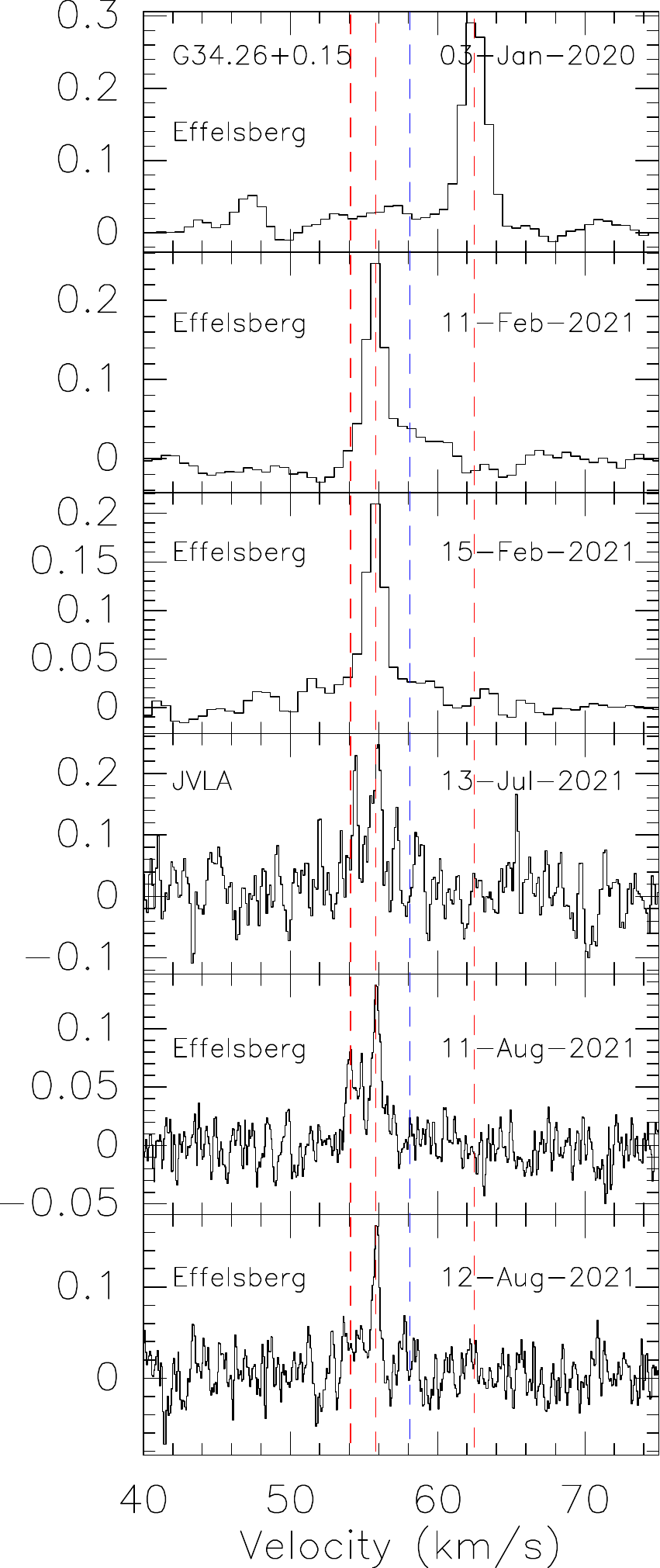}
  \caption{Spectra from NH$_3$ (9,6) transition lines. \textbf{Left}: \textit{Top to bottom:} Time sequence of NH$_3$ (9,6) profiles observed toward Cep A with the Effelsberg 100-meter telescope (after subtracting a first-order polynomial baseline). A JVLA spectrum is interspersed. The systemic velocity from CO and HCO$^+$ lines is indicated by a dashed blue line. The two dashed red lines at LSR velocities, $V_{\rm LSR}$, of $-$0.90~km~s$^{-1}$ and $-$0.28~km~s$^{-1}$ indicate the central velocities of the two major components. \textbf{Right}: NH$_3$ (9,6) spectra from G34.26$+$0.15. The systemic velocity from C$^{17}$O is indicated by a dashed blue line. The three dashed red lines at $V_{\rm LSR}$~=~54.1~km~s$^{-1}$, 55.8~km~s$^{-1}$, and 62.5~km~s$^{-1}$ show the central velocities of the main ammonia emission components.}
  \label{spectra_all}
\end{figure}

\begin{figure}[h]
\center
        \includegraphics[height=152pt]{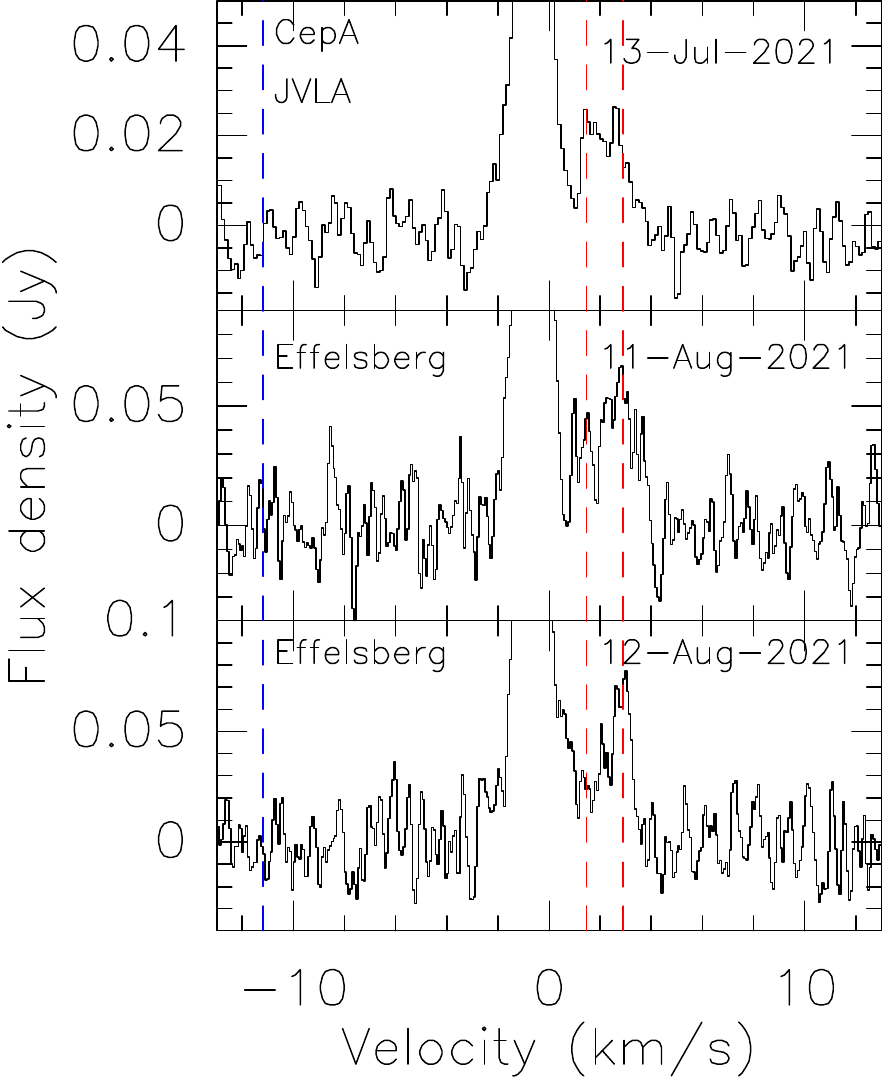}
        \includegraphics[height=152pt]{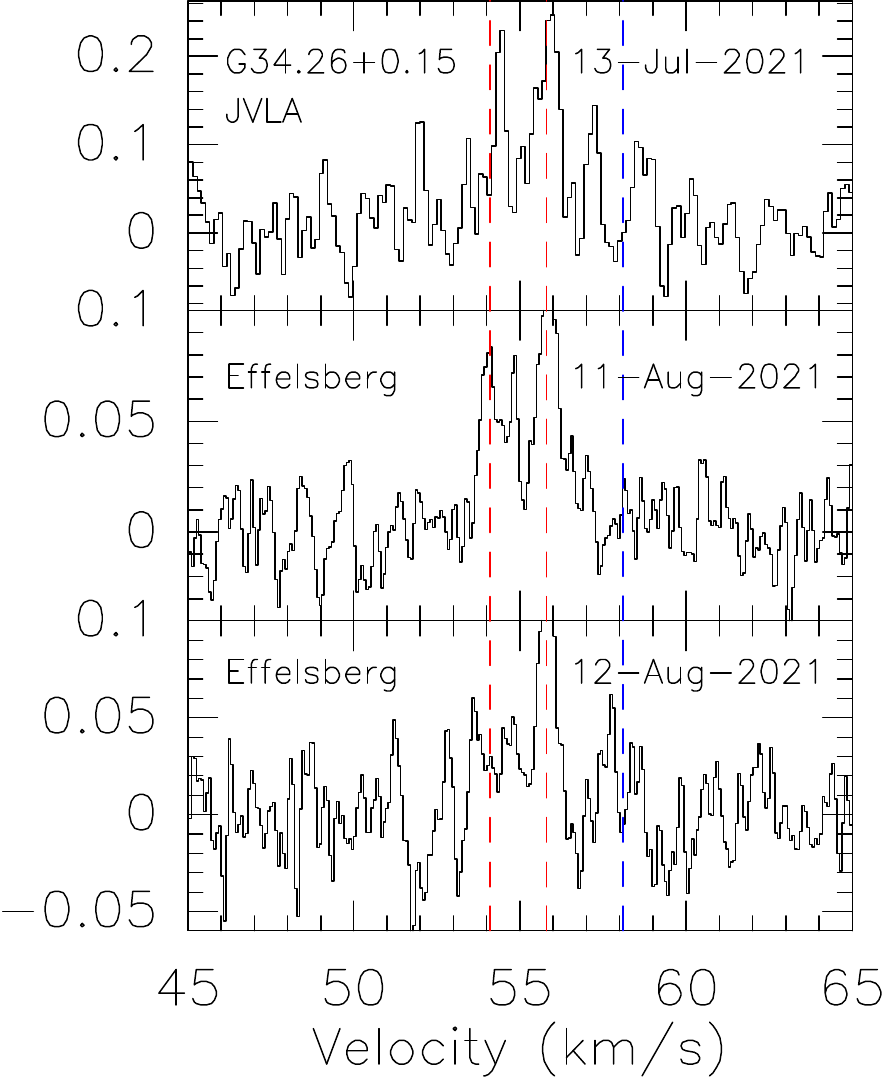}
  \caption{NH$_3$ (9,6) line profiles emphasizing, in contrast to the spectra in Fig. \ref{spectra_all}, weaker features. Cep A spectra are presented on the left, G34.26$+$0.15 spectra on the right. The two dashed red lines in the left panels indicate $V_{\rm LSR}$~=~1.48 km~s$^{-1}$ and 2.89~km~s$^{-1}$. In the right panels, the two dashed red lines refer to 54.1~km~s$^{-1}$ and 55.8~km~s$^{-1}$.}
  \label{spectra_zoom}
\end{figure}

\section{Results}
\label{results}

The spectra from different epochs are shown in Figs.~\ref{spectra_all} and \ref{spectra_zoom}. Toward Cep A, the NH$_3$ (9,6) line profile from the JVLA is extracted from an Effelsberg-beam-sized region (FWHM, 49$\arcsec$). In the case of G34.26$+$0.15, the NH$_3$ spectrum is below the noise level if a similarly large beam size is used. Therefore, we derived the JVLA NH$_3$ (9,6) spectrum from a smaller region, with radius 3$\farcs$5, that contains all the detected NH$_3$ (9,6) emission. In Table~\ref{spectra_fitting}, the observed NH$_3$ (9,6) line parameters obtained by Gaussian fits are listed. NH$_3$ (8,5) and (10,7) emission is not detected by our JVLA observations. The 3$\sigma$ upper limits for the NH$_3$ (8,5) and (10,7) lines toward Cep A are 23.2 mJy~beam$^{-1}$ and 27.2~mJy~beam$^{-1}$, respectively. In G34.26$+$0.15, the corresponding 3$\sigma$ upper limits for the NH$_3$ (8,5) and (10,7) lines are 22.1~mJy~beam$^{-1}$ and 30.4~mJy~beam$^{-1}$. For both sources, sensitivity levels refer to emission from a single channel of width 0.13~km~s$^{-1}$. Taking the larger measured line widths of the (9,6) maser features (see Table \ref{spectra_fitting}), these limits could be further lowered by factors of two to four.

\subsection{Centimeter-continuum emission}

\begin{figure*}[h]
\center
        \includegraphics[width=500pt]{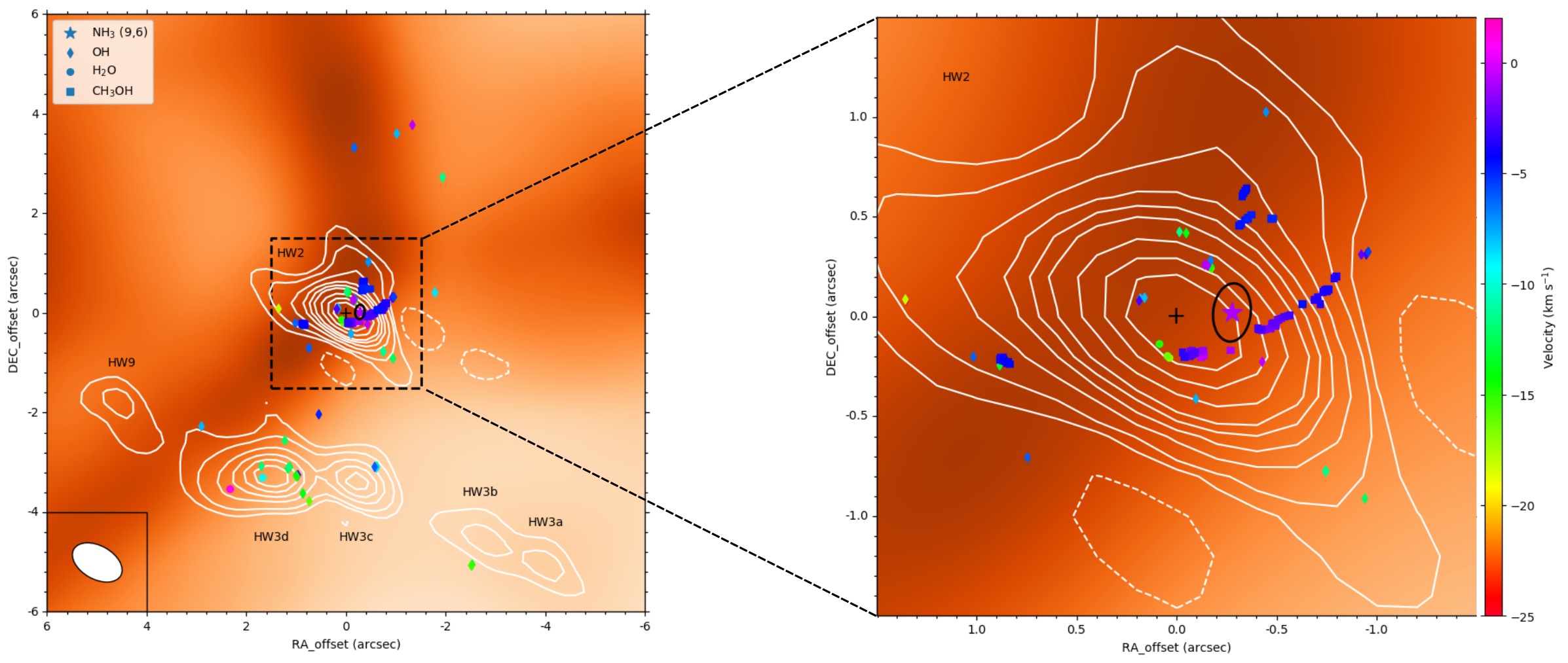}
  \caption{Cepheus A. White contours mark the 1.36\,cm JVLA continuum map of Cep A; levels are $-$5, 5, 10, 20, 30, 40, 50, 70, 90, and 110~$\times$~0.125~mJy~beam$^{-1}$. The background image is the \textit{Spitzer} 4.5$\,\mu\rm{m}$ emission, taken from the Galactic Legacy Infrared Mid-Plane Survey Extraordinaire \citep[GLIMPSE;][]{2003PASP..115..953B,2009PASP..121..213C}. The reference position is $\alpha_{\rm J2000}$~=~22$^{\rm h}$56$^{\rm m}$17$\fs$972, and $\delta_{\rm J2000}$~=~62$\degr$01$\arcmin$49$\farcs$587, the peak position of the continuum map, is marked with a black cross. Slightly to the west of the cross is the black ellipse denoting the position of the NH$_3$ (9,6) emission with a purple star at its center. OH \citep{2005MNRAS.361..623B}, H$_2$O \citep{2018ApJ...856...60S}, and CH$_3$OH \citep{2017A&A...603A..94S} masers are presented as diamonds, circles, and squares, respectively. The color bar on the right-hand side indicates the LSR velocity range of the maser spots.}
  \label{cepa_allcontinuum}
\end{figure*}

\begin{figure*}[h]
\center
        \includegraphics[width=500pt]{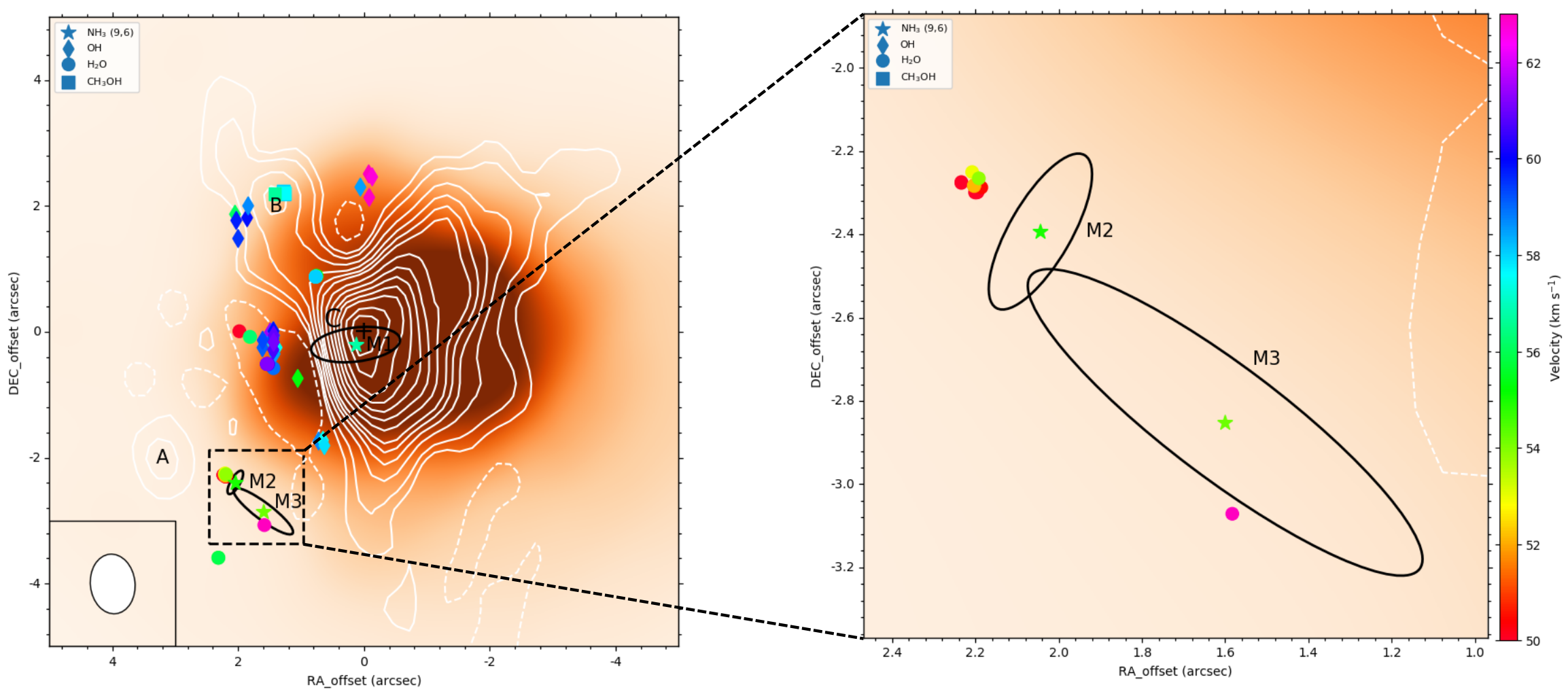}
  \caption{1.36\,cm JVLA continuum map of G34.26$+$0.15 presented as white contours with levels of $-$5, 5, 10, 20, 30, 40, 50, 70, 90, 110, 130, 150, 180, and 200~$\times$~5.0~mJy~beam$^{-1}$. The background image is the \textit{Spitzer} 4.5$\,\mu\rm{m}$ emission, taken from GLIMPSE. The reference position is $\alpha_{\rm J2000}$~=~18$^{\rm h}$53$^{\rm m}$18$\fs$560, and $\delta_{\rm J2000}$~=~01$\degr$14$\arcmin$58$\farcs$201, the peak position, is marked by a black cross. The black ellipses show the positions of NH$_3$ (9,6) emissions with stars at their center (i.e., M1, M2, and M3). OH \citep{2000MNRAS.317..192Z}, H$_2$O \citep{2011PASJ...63.1293I}, and CH$_3$OH \citep{2016A&A...587A.104B} masers are presented as diamonds, circles, and squares, respectively. The color bar indicates the velocity range ($V_{\rm LSR}$) of maser spots.}
    \label{g34_allcontinuum}
\end{figure*}

The 1.36\,cm continuum, derived from our JVLA observations, toward Cep A is presented in Fig. \ref{cepa_allcontinuum}. Six published compact sources, HW2, HW3a, HW3b, HW3c, HW3d, and HW9, are detected in our observations. Figure~\ref{g34_allcontinuum} shows the 1.36\,cm continuum in G34.26$+$0.15. Three main continuum objects, A, B, and C, are detected. By using the \textit{imfit} task in CASA, we measured the continuum flux at 1.36\,cm toward individual compact source components in Cep A and G34.26$+$0.15. Details are given in Table~\ref{continuum_sou}.

\subsection{NH$_3$ (9,6) emission in Cep A}
In 2020 January, NH$_3$ (9,6) emission with a peak flux density of 0.67~$\pm$~0.07~Jy was first detected with the Effelsberg 100-meter telescope in Cep A. Emission with similar strength was also detected in 2021 February and August with the same telescope. Higher velocity resolution data, which were obtained in 2021 August, again with the Effelsberg 100-meter telescope, show that the (9,6) emission contains two main velocity components. Overall, the flux densities of the NH$_3$ (9,6) emission line measured with the Effelsberg 100-meter telescope are, within the calibration uncertainties, unchanged. This is valid for the time interval between 2020 January and August 2021, when we smoothed the obtained spectra to the same velocity resolution. We also see another two weaker components. Figure~\ref{spectra_zoom} emphasizes these weak components with an expanded flux density scale. 

Higher angular resolution data from the JVLA pinpoint the position of the NH$_3$ (9,6) emission with an offset of ($-0\farcs$28, $0\farcs$02) relative to the 1.36\,cm continuum peak of Cep~A~HW2 (Fig. \ref{cepa_allcontinuum}). The deconvolved NH$_3$ (9,6) component size is $(0\farcs29\pm0\farcs15) \times (0\farcs19\pm0\farcs14)$ at P.A.~=~$174\degr$, derived with the \textit{imfit} task in CASA, and can thus be considered, accounting for the uncertainties, as unresolved.

In view of the constancy of the flux densities obtained at Effelsberg and the similar JVLA flux density, measured in 2021 July, there is no missing interferometric flux density in the JVLA data.

\subsection{NH$_3$ (9,6) emission in G34.26$+$0.15}

The NH$_3$ (9,6) emission was first detected toward G34.26$+$0.15 in 2020 January with the Effelsberg 100-meter telescope. Higher velocity resolution data from 2021 August show the NH$_3$ (9,6) emission to be composed of two different components. The spectra of weak components on a smaller flux density scale are presented in Fig.~\ref{spectra_zoom}.

Three different locations showing NH$_3$ (9,6) emission are found toward G34.26$+$0.15 (Fig.~\ref{g34_allcontinuum}). The deconvolved NH$_3$ (9,6) component sizes are $(1\farcs42\pm0\farcs43) \times (0\farcs54\pm0\farcs62)$ at P.A.~=~$97\degr$ (M1), $(0\farcs42\pm0\farcs27) \times (0\farcs15\pm0\farcs27)$ at P.A.~=~$150\degr$ (M2), and $(1\farcs17\pm0\farcs34) \times (0\farcs27\pm0\farcs46)$ at P.A.~=~$53\degr$ (M3) and are thus comparable to or smaller than the beam size. 

Overall, the NH$_3$ (9,6) line from G34.26$+$0.15 weakened during the time interval from 2020 January to 2021 August by about 70\%. A comparison between the JVLA spectrum and the Effelsberg data, assuming a linear decrease in the integrated intensity as a function of time between different epochs of the 100-meter observations, suggests there is no missing flux in the JVLA data. This is similar to the situation in Cep A.

\section{Discussion}
\label{disscussion}
\subsection{Morphology of Cep A and G34.26$+$0.15}
\label{morphology}

Cep A, at a trigonometric parallax distance of 0.70$\pm$0.04~kpc \citep{2009ApJ...693..406M,2011ApJ...733...71D}, is the second closest HMSFR (after Orion) and by far the closest NH$_3$ (9,6) maser known. About 16 compact ($\sim$1$\arcsec$) radio sources \citep[e.g.,][]{1984ApJ...276..204H,1991ApJ...383..280H,1996ApJ...459..193G} have been identified in Cep A. \citet{1984ApJ...276..204H} discovered these targets at radio wavelengths, which are UC and hypercompact (HC) \h2 regions and/or stellar wind sources,  subsequently named as HW sources. The HW2 object is one of the best known examples of a protostellar jet or disk system driving a powerful outflow \citep[e.g.,][]{1980ApJ...240L.149R,1984A&A...138..205G,1986ApJ...305..721T,2006ApJ...638..878C,2021ApJ...914L...1C}. The observed NH$_3$ (9,6) emission is slightly offset ($-0\farcs$28, $0\farcs$02) from the center of HW2 (see Fig.~\ref{cepa_allcontinuum}).

G34.26$+$0.15 is an HMSFR located at a distance of 3.3 kpc \citep{1994ApJ...436..117K}. It hosts four radio continuum components named A, B, C, and D. Component C is a prototypical cometary UC \h2 region containing a compact head and a diffuse tail that extends from east to west \citep[e.g.,][]{1985ApJ...288L..17R,1986ApJ...309..553G,2004ApJ...605..285S,2011ApJS..194...44S}. Components A and B are HC \h2 regions, located to the east of component C. An extended ring-like \h2 region, called component D, is located southeast of components A-C. One of the three observed NH$_3$ (9,6) emission line sources, M1, is close to the head of component C, whereas M2 and M3 originate from another compact region in the west of the HC \h2 component A (see  Fig.~\ref{g34_allcontinuum}).

\subsection{NH$_3$ (9,6) emission possibly caused by maser action}

As shown in Fig. \ref{spectra_all}, the NH$_3$ (9,6) profiles in Cep A and G34.26$+$0.15 are narrow ($\Delta V_{1/2}\le$2.0~km~s$^{-1}$), much narrower than the expected line widths ($\gtrsim$4~km~s$^{-1}$) of thermal lines observed at a similar angular resolution  \citep[e.g.,][]{1985ApJ...288..595T,1986ApJ...305..721T,1993ApJ...410..202T,1999MNRAS.307...58T,1987A&A...182..137H,2007A&A...469..207C,2007ApJ...659..447M,2012A&A...542L..15W,2018A&A...617A.100B}. Velocity shifts with respect to the systemic velocities of the two sources are both observed, that is, $ V\sim$10~km~s$^{-1}$ in Cep A and $ V\sim$4~km~s$^{-1}$ in G34.26$+$0.15 (see details in Sect. \ref{nh3_thermal}). Furthermore, time variability is observed in the case of G34.26$+$0.15, which is also a characteristic feature of maser emission. 

Additional evidence of their maser nature is the high brightness temperatures of the (9,6) emission spots toward Cep A and G34.26$+$0.15. The spectral parameters are listed in Table \ref{96positions}. Because at least a significant part of the NH$_{3}$ (9,6) emission is not resolved by our JVLA observations, the derived brightness temperatures are only lower limits. Nevertheless, the lower limits on the brightness temperature are $>$800~K in Cep A (see Table~\ref{96positions}), which is much higher than the expected thermal gas temperature of $\sim$250~K \citep[e.g.,][]{2005Natur.437..109P,2007A&A...469..207C,2018A&A...617A.100B}. This strongly suggests that the NH$_3$ (9,6) emission in Cep A is due to maser action. Because G34.26+0.15 is located at about five times the distance to Cep A, beam dilution effects reduce the lower main beam brightness temperature limit to 400 K in G34.26$+$0.15 (M2) (see Table~\ref{96positions}). We also note that the luminosity of the NH$_3$ (9,6) emission in G34.26$+$0.15 is higher than or comparable to that in Cep A, depending on the epoch of our observations. 

Finally, the non-detections of the (8,5) and (10,7) lines also indicate that the (9,6) line is special. This allows us to derive lower 3$\sigma$ limits of the (9,6)/(8,5) and (9,6)/(10,7) line intensity ratios. The (9,6) line arises from ortho-NH$_3$ ($K=3n$), whereas the NH$_3$ (8,5) and (10,7) lines are para-NH$_3$ ($K\neq3n$) lines. The minimum ortho-to-para ratios are in the range 12--42 and 1--8 toward Cep A and G34.26$+$0.15, respectively. The statistical weights for the ortho states are twice as large as those for the para states \citep[e.g.,][]{1999ApJ...525L.105U,2011ApJ...739L..13G,2013A&A...549A..90H}. In Cep A, the line intensity ratios are far higher than this factor of two. Thus, at least in Cep A the higher main beam brightness peak temperature of the (9,6) emission is caused by maser action, perhaps involving exponential amplification, and the case of G34.26$+$0.15 is likely similar.

\subsection{Comparison of NH$_3$ (9,6) masers with previously published (quasi-)thermal NH$_3$ emission}
\label{nh3_thermal}

The metastable (1,1), (2,2), (3,3), and (4,4) ammonia lines show thermal emission toward Cep A over a velocity range of $-$13~km~s$^{-1}\leq V_{\rm LSR}~\leq-$4~km~s$^{-1}$ \citep{1981MNRAS.195..607B,1984A&A...138..205G,1985ApJ...288..595T,1986ApJ...305..721T,1993ApJ...410..202T,1999MNRAS.307...58T}. An average NH$_3$ column density of $\sim$5$\times$10$^{15}$~cm$^{-2}$ was estimated for a region of 3\arcsec\,around HW2 \citep{1999MNRAS.307...58T}. This high NH$_3$ abundance could provide a suitable environment for maser species. Large line widths ($\Delta V_{1/2}\simeq$7.0~km~s$^{-1}$) with $V_{\rm LSR}\sim~-$10~km~s$^{-1}$ in both (1,1) and (2,2) lines were found toward HW2 \citep{1993ApJ...410..202T}. The velocity is similar to the cloud's systemic local standard of rest (LSR) velocity of $-11.2$~km~s$^{-1}$, which is based on CO \citep{1996ApJ...466..844N} and HCO$^+$ observations \citep{1999ApJ...514..287G}. Our (9,6) maser is redshifted ($-$0.9~km~s$^{-1}\leq V_{\rm LSR}~\leq$2.9~km~s$^{-1}$) and shares positions with the outflowing gas seen in CO and HCO$^+$ with similarly redshifted velocities. Therefore, we argue that the (9,6) masers are related to outflowing gas.

In G34.26$+$0.15, a large NH$_3$ column density, 10$^{18.5\pm0.2}$~cm$^{-2}$, and a kinetic temperature of 225$\pm$75 K were derived by \citet{1987A&A...182..137H} based on measurements of 15 NH$_3$ inversion transitions in the frequency range of 22.0--26.0 GHz. These did not include the (9,6) transition. While these lines were measured with a beam size of about 40$\arcsec$, a comparison of the peak intensities of the optically thick lines with the kinetic temperature reveals the size of the hot, ammonia-emitting core to be only $\sim$2.5$\arcsec$. All those measured NH$_3$ lines were quasi-thermal and had LSR velocities of $\sim~$58.5~km~s$^{-1}$, close to the systemic velocity of $\sim~$58.1~km~s$^{-1}$ obtained from C$^{17}$O observations \citep{2012A&A...542L..15W}. Their line widths ($\Delta V_{1/2}\geq$3.6~km~s$^{-1}$) are larger than what we find (0.35~km~s$^{-1}\leq\Delta V_{1/2}\leq$~0.94~km~s$^{-1}$) for each (9,6) maser component (see details in Table \ref{96positions}). In all, we may have observed four different (9,6) velocity features. Three are blueshifted at $V_{\rm LSR}~\sim~$53.8~km~s$^{-1}$, 55.8~km~s$^{-1}$, and 56.8~km~s$^{-1}$, and a fourth, tentatively detected, at 62.5 ~km~s$^{-1}$. This tentative redshifted feature was only potentially detected with Effelsberg in 2020 January. The velocity is similar to that of the JVLA measurements on the NH$_3$ (1,1) absorption line against continuum source C \citep[$\sim7\arcsec$ resolution;][]{1987ApJ...323L.117K} and the NH$_3$ (3,3) emission surrounding continuum source B as well as the head of C \citep[1$\farcs$4$\times$1$\farcs$2 resolution;][]{1989A&A...213..148H}. However, we did not find this redshifted component in our JVLA observations. Therefore, its position within G34.26$+$0.15 cannot be determined. The blueshifted (9,6) masers with a velocity range of 53.8--56.8~km~s$^{-1}$ (M1, M2, and M3) show velocities compatible with those of the NH$_3$ (3,3) emission at the proper positions \citep{1989A&A...213..148H}, which might be a suitable environment for maser species.

\subsection{Comparison of NH$_3$ (9,6) masers with other maser lines}

To characterize the environment of NH$_3$ (9,6) masers, we can compare their positions with respect to those of other maser species (i.e., OH, H$_2$O, and CH$_3$OH). Toward Cep A HW2, many CH$_3$OH \citep[e.g.,][]{1991ApJ...380L..75M,2008PASJ...60.1001S,2017A&A...603A..94S} and H$_2$O maser spots \citep[e.g.,][]{1998ApJ...509..262T,2011MNRAS.410..627T,2018ApJ...856...60S} are detected and are associated with its disk. \citet{2018ApJ...856...60S} also found that most of the H$_2$O maser flux is associated with the compact \h2 region HW3d. OH maser features close to the \h2 regions are also seen in HW2 \citep[e.g.,][]{1985MNRAS.216P..51C,2005MNRAS.361..623B}. These three kinds of masers in Cep A have a large velocity range of $-$25~km~s$^{-1}\leq V_{\rm LSR}~\leq-$2~km~s$^{-1}$ and are widespread around HW2 and HW3, while NH$_3$ (9,6) emission is only detected at $-$0.9~km~s$^{-1}\leq V_{\rm LSR}~\leq$2.9~km~s$^{-1}$ toward a sub-arcsecond-sized region to the west of the peak continuum position of HW2 (see Fig.~\ref{cepa_allcontinuum}). This suggests that the NH$_3$ (9,6) maser in Cep A is unique and not related to maser spots seen in other molecular species.

In G34.26$+$0.15, OH \citep{2000MNRAS.317..192Z}, H$_2$O \citep{2011PASJ...63.1293I}, and CH$_3$OH \citep{2016A&A...587A.104B} masers have been detected east of source C (Fig.~\ref{g34_allcontinuum}), and none of them coincides with the head of C. The NH$_3$ (9,6) maser M1 is also found slightly off the head of source C. This could suggest that M1 is powered by continuum source C or by an outflow. Near component B, there are some OH and CH$_3$OH masers but no H$_2$O or NH$_3$ masers. A group of H$_2$O masers, well-known tracers of outflows, with a large velocity distribution of 43~km~s$^{-1}\leq V_{\rm LSR}~\leq$54~km~s$^{-1}$, was found to the west of the centimeter-continuum source A and close to the peak of the millimeter-continuum emission (see details in our Fig.~\ref{g34_continuum_cm_mm} and also in Fig.~5 of \citealt{2011PASJ...63.1293I}). The closeness of NH$_3$ (9,6) maser spots M2 and M3 to this group of water masers and their similar velocities again suggest an association of NH$_3$ (9,6) masers with outflow activity.

\subsection{Constraints on pumping scenarios}
Our observations have resulted in the detection of NH$_{3}$ (9,6) masers in Cep A and G34.26$+$0.15. The new detections could provide additional constraints on the maser line's pumping mechanism. As mentioned in Sect.~\ref{introduction}, the pumping mechanism of the (9,6) maser is unclear \citep{1986ApJ...300L..79M,1991ApJ...378..445B}. Previous studies have suggested that there are three main pumping scenarios to explain the observed NH$_3$ maser lines \citep{1986ApJ...300L..79M,2013A&A...549A..90H}: (1) infrared radiation from the dust continuum emission, (2) line overlap, and (3) collisional pumping. 

For the first mechanism, infrared photons near 10~$\mu$m are needed for vibrational excitation. The high dust temperature ($\sim$300~K) of W51-IRS2 can provide substantial infrared photons near 10~$\mu$m, which is used for radiative pumping \citep{2013A&A...549A..90H}. Both Cep A and G34.26$+$0.15 have similar kinetic temperatures of $\gtrsim$200~K \citep{1987A&A...182..137H,2005Natur.437..109P,2007A&A...469..207C,2018A&A...617A.100B}. This suggests that high kinetic temperatures are needed to excite NH$_{3}$ (9,6) masers. However, it should be noted that the silicate dust absorption feature might dominate at 10~$\mu$m (see the spectral energy distribution of Cep A in \citealt{2017ApJ...843...33D}). Additionally, there is no bright infrared emission around the two (9,6) masers, M2 and M3, in G34.26$+$0.15 (see Fig.~\ref{g34_allcontinuum}; see also Fig.~11 in \citealt{2003ApJ...598.1127D} for a 10.5~$\mu$m map). This indicates that the pumping mechanism via infrared photons near 10~$\mu$m may not be viable to explain the (9,6) masers in Cep A and G34.26$+$0.15. Furthermore, \citet{1993LNP...412..123W} argued that radiative pumping by dust emission tends to excite multiple adjacent ammonia maser transitions, which appears to contradict our failure to detect the adjacent (8,5) and (10,7) lines (with respect to quantum numbers and frequency) and to only measure the (9,6) transitions in Cep A and G34.26$+$0.15. Therefore, we suggest that infrared radiation from dust is not the main pumping source.

\citet{1986ApJ...300L..79M} suggested that there might be some line overlaps between the rotational NH$_{3}$ transitions in the far-infrared band. However, this would be unlikely to affect only the (9,6) line. Nevertheless, far-infrared spectral observations will be needed to clarify this scenario.

Based on our observations, the (9,6) maser spots are close to, but not coincident with, the peaks of the radio continuum emission in Cep A and G34.26$+$0.15. Furthermore, the (9,6) masers show velocity offsets with respect to their systemic velocities. This indicates that the (9,6) masers are located at the base of outflows, similar to the H$_{2}$O masers. This is supported by VLBI observations that show that (9,6) masers tend to be closely associated with H$_{2}$O masers \citep{1991ApJ...373L..13P}. The observed time variability in G34.26$+$0.15 and W51-IRS2 can also be attributed to episodic molecular outflows. This indicates that collisional pumping could be the driver of the (9,6) maser. On the other hand, collisional pumping has been successfully used to explain the NH$_{3}$ (3,3) maser \citep{1983A&A...122..164W,1990MNRAS.244P...4F,1994ApJ...428L..33M}. Collisions tend to pump from the $K$=0 level to the $K$=3 level with parity changes, that is, the upper level of the (3,3) metastable transition will be overpopulated. NH$_{3}$ (9,6) arises from the ortho species, so a similar mechanism might also occur in the case of the (9,6) transition. Further measurements of collisional rates of ammonia will allow us to test this scenario.

\section{Summary}
\label{summary}

We report the discovery of NH$_3$ (9,6) masers in two HMSFRs, Cep A and G34.26$+$0.15. The narrow line width of the emission features ($\Delta V_{1/2}\le$2.0~km~s$^{-1}$) and their high brightness temperatures ($>400$~ K) indicate the maser nature of the lines. The intensity of the (9,6) maser in G34.26$+$0.15 is decreasing with time, while toward Cep A the maser is stable based on 20 months of monitoring at Effelsberg. Linearly interpolating the integrated intensities obtained at Effelsberg as a function of time, the JVLA measurements show that there is no missing flux density on scales on the order of 1.2 arcsec (4~$\times 10^{-3}$ and 2~$\times 10^{-2}$\,pc) to the total single-dish flux. The JVLA-detected emission indicates that the NH$_{3}$ (9,6) maser in Cep A originates from a sub-arcsecond-sized region slightly ($0\farcs28 \pm 0\farcs10$) to the west of the peak position of the 1.36\,cm continuum object, HW2. In G34.26$+$0.15, three NH$_{3}$ (9,6) maser spots are observed: one is close to the head of the cometary UC \h2 region C, and the other two are emitted from a compact region to the west of the HC \h2 region A. We suggest that the (9,6) masers may be connected to outflowing gas. Higher angular resolution JVLA and VLBI observations are planned to provide more accurate positions and constraints on pumping scenarios.

\begin{acknowledgements}
We would like to thank the anonymous referee for the useful comments that improve the manuscript. Y.T.Y. is a member of the International Max Planck Research School (IMPRS) for Astronomy and Astrophysics at the Universities of Bonn and Cologne. Y.T.Y. would like to thank the China Scholarship Council (CSC) for its support. We would like to thank the staff at the Effelsberg for their help provided during the observations. We thank the staff of the JVLA, especially Tony Perreault and Edward Starr, for their assistance with the observations and data reduction. This research has made use of the NASA/IPAC Infrared Science Archive, which is funded by the National Aeronautics and Space Administration and operated by the California Institute of Technology.
\end{acknowledgements}

\bibliographystyle{aa}
\bibliography{mainArxiv}

\begin{appendix}
\onecolumn
\section{}
\begin{table}[h]
\caption{Summary of NH$_3$ (9, 6) maser observations.}
\centering
\begin{tabular}{lccccccccc}
\hline\hline
%\tablenum{2}
Source & Telescope & Beam & Epoch &  Channel  & $S_\nu$  & rms & $\int S_\nu dv$ & $V_{\rm LSR}$ & $\Delta V_{1/2}$  \\
  &   & size & & spacing  &    &  & & &   \\
  &   &  & & (km s$^{-1}$) & (Jy) &  (mJy)  & (Jy km s$^{-1}$) & \multicolumn{2}{c}{ (km s$^{-1}$)}    \\
\hline
\label{spectra_fitting}
Cep A      & Effelsberg & 49$\arcsec$ & 2020, Jan. 04  & 0.62 & 0.67 & 3.41  & 1.19 $\pm$ 0.02 & -1.11 $\pm$ 0.02 & 1.67 $\pm$ 0.04   \\
           & Effelsberg & 49$\arcsec$ & 2021, Feb. 11  & 0.62 & 0.59 & 5.97  & 1.08 $\pm$ 0.02 & -0.74 $\pm$ 0.02 & 1.70 $\pm$ 0.04  \\
           & Effelsberg & 49$\arcsec$ & 2021, Feb. 15  & 0.62 & 0.65 & 10.98 & 1.11 $\pm$ 0.03 & -0.75 $\pm$ 0.02 & 1.60 $\pm$ 0.05 \\
  & JVLA\tablefootmark{a}& $1\farcs47~\times~0\farcs99$ & 2021, Jul. 13  & 0.13 & 1.13 & 144  & 0.89 $\pm$ 0.09 & -0.86 $\pm$ 0.03 & 0.74 $\pm$ 0.12 \\
           & Effelsberg & 49$\arcsec$ & 2021, Aug. 11  & 0.07 & 0.98 & 13.36 & 0.49 $\pm$ 0.02 & -0.90 $\pm$ 0.01 & 0.47 $\pm$ 0.01 \\
           &            &             &                &      & 0.35 &       & 0.26 $\pm$ 0.02 & -0.28 $\pm$ 0.02 & 0.69 $\pm$ 0.05 \\
           & Effelsberg & 49$\arcsec$ & 2021, Aug. 12  & 0.07 & 0.98 & 13.35 & 0.50 $\pm$ 0.01 & -0.89 $\pm$ 0.07 & 0.48 $\pm$ 0.07 \\
           &            &             &                &      & 0.35 &       & 0.20 $\pm$ 0.01 & -0.29 $\pm$ 0.07 & 0.54 $\pm$ 0.07 \\
           &            &             &                &      & 0.06 &       & 0.07 $\pm$ 0.01 &  0.51 $\pm$ 0.07 & 1.09 $\pm$ 0.07 \\
           &            &             &                &      & 0.02 &       & 0.02 $\pm$ 0.01 &  2.15 $\pm$ 0.07 & 0.80 $\pm$ 0.07 \\
           &            &             &                &      & 0.07 &       & 0.06 $\pm$ 0.01 &  2.89 $\pm$ 0.07 & 0.92 $\pm$ 0.07 \\
G34.26+0.15& Effelsberg & 49$\arcsec$ & 2020, Jan. 03  & 0.62 & 0.30 & 1.26  & 0.65 $\pm$ 0.03 & 62.50 $\pm$ 0.05 & 2.05 $\pm$ 0.13 \\
           & Effelsberg & 49$\arcsec$ & 2021, Feb. 11  & 0.62 & 0.24 & 2.42  & 0.40 $\pm$ 0.02 & 55.76 $\pm$ 0.04 & 1.60 $\pm$ 0.12 \\
           & Effelsberg & 49$\arcsec$ & 2021, Feb. 15  & 0.62 & 0.20 & 4.86  & 0.38 $\pm$ 0.02 & 55.71 $\pm$ 0.05 & 1.80 $\pm$ 0.14 \\
 &JVLA\tablefootmark{b}& $1\farcs33~\times~1\farcs06$ & 2021, Jul. 13  & 0.13 & 0.23 & 37.1  & 0.09 $\pm$ 0.02 & 54.41 $\pm$ 0.03 & 0.38 $\pm$ 0.09 \\
           &            &             &                &      & 0.22 &       & 0.22 $\pm$ 0.02 & 55.82 $\pm$ 0.05 & 0.95 $\pm$ 0.12 \\
           &            &             &                &      & 0.15 &       & 0.06 $\pm$ 0.01 & 57.21 $\pm$ 0.04 & 0.35 $\pm$ 0.08 \\
           & Effelsberg & 49$\arcsec$ & 2021, Aug. 11  & 0.07 & 0.08 & 13.92 & 0.06 $\pm$ 0.007 & 54.10 $\pm$ 0.05 & 0.68 $\pm$ 0.12 \\
           &            &             &                &      & 0.07 &       & 0.02 $\pm$ 0.006 & 54.82 $\pm$ 0.03 & 0.31 $\pm$ 0.09 \\
           &            &             &                &      & 0.12 &       & 0.10 $\pm$ 0.006 & 55.85 $\pm$ 0.02 & 0.75 $\pm$ 0.06 \\
           & Effelsberg & 49$\arcsec$ & 2021, Aug. 12  & 0.07 & 0.16 & 27.40 & 0.09 $\pm$ 0.008 & 55.83 $\pm$ 0.02 & 0.56 $\pm$ 0.05 \\
\hline
\end{tabular}
\tablefoot{
The spectral parameters are obtained from Gaussian fitting.
\tablefoottext{a}{The JVLA spectrum toward Cep A is extracted from the Effelsberg-beam-sized region (FWHM 49$\arcsec$).} 
\tablefoottext{b}{For G34.26$+$0.15, the JVLA beam samples the NH$_3$ (9,6) spectrum over a region of radius  3$\farcs$5, which contains all detected NH$_3$ (9,6) emissions.} }
\end{table}

\begin{table}[h]
\caption{1.36 cm JVLA flux densities of individual continuum sources.}
\centering
\begin{tabular}{lcccccc}
\hline\hline
%\tablenum{2}
\multicolumn{2}{c}{Source} &  R.A. & Dec.  & Size  & P.A.  &    $S_\nu$    \\
 & & ($h\quad m\quad s$) & ($\degr\quad \arcmin\quad \arcsec$) & (arcsec)  & (deg) & (mJy) \\
\hline
\label{continuum_sou}
Cep A      & HW2  & 22 56 17.972 $\pm$ 0.003 & $+$62 01 49.587 $\pm$ 0.015 & (0.45 $\pm$ 0.19) $\times$ (0.22 $\pm$ 0.10)   & 50.0   & 20.2 $\pm$ 1.4 \\
           & HW3a & 22 56 17.420 $\pm$ 0.022 & $+$62 01 44.576 $\pm$ 0.076 & (2.35 $\pm$ 0.45) $\times$ (0.55 $\pm$ 0.14)   & 66.6   & 4.75 $\pm$ 0.74\\
           & HW3b & 22 56 17.578 $\pm$ 0.009 & $+$62 01 45.041 $\pm$ 0.043 & (1.43 $\pm$ 0.24) $\times$ (0.45 $\pm$ 0.10)   & 59.9   & 3.19 $\pm$ 0.36\\
           & HW3c & 22 56 17.956 $\pm$ 0.016 & $+$62 01 46.224 $\pm$ 0.038 & (1.44 $\pm$ 0.37) $\times$ (0.36 $\pm$ 0.19)   & 86.0   & 9.90 $\pm$ 1.7  \\
           & HW3d & 22 56 18.195 $\pm$ 0.005 & $+$62 01 46.325 $\pm$ 0.014 & (1.26 $\pm$ 0.12) $\times$ (0.30 $\pm$ 0.19)   & 102.5  & 13.75 $\pm$ 0.92\\
           & HW9  & 22 56 18.626 $\pm$ 0.014 & $+$62 01 47.851 $\pm$ 0.137 & (1.53 $\pm$ 0.51) $\times$ (0.29 $\pm$ 0.30)   & 28.0   & 3.26 $\pm$ 0.78 \\
G34.26+0.15& A    & 18 53 18.774 $\pm$ 0.005 & $+$01 14 56.208 $\pm$ 0.125 & (0.66 $\pm$ 0.49) $\times$ (0.50 $\pm$ 0.33)   & 10.0   & 94 $\pm$ 33     \\
           & B    & 18 53 18.649 $\pm$ 0.005 & $+$01 15 00.071 $\pm$ 0.180 & (2.31 $\pm$ 0.49) $\times$ (0.85 $\pm$ 0.21)   & 17.4   & 597 $\pm$ 110   \\
           & C    & 18 53 18.560 $\pm$ 0.004 & $+$01 14 58.201 $\pm$ 0.112 & (2.03 $\pm$ 0.30) $\times$ (1.34 $\pm$ 0.20)   & 178.0  & 5070 $\pm$ 660  \\
\hline
\end{tabular}
%\tablefoot{}
\end{table}

\begin{table}[h]
\caption{NH$_3$ (9,6) maser positions derived from the JVLA observations.}
\centering
\begin{tabular}{llcccccc}
\hline\hline
%\tablenum{2}
\multicolumn{2}{c}{Source} & R.A. & Dec. &  $S_\nu$  & $T_{\rm MB}$  & $V_{\rm LSR}$ & $\Delta V_{1/2}$  \\
 & &   &  &    &    &   &   \\
 & & ($h\quad m\quad s$) & ($\degr\quad \arcmin\quad \arcsec$) & (mJy~beam$^{-1}$) & (K) &   \multicolumn{2}{c}{ (km~s$^{-1}$)}    \\
\hline
\label{96positions}
Cep A      & M  & 22 56 17.933 $\pm$ 0.002 & $+$62 01 49.608 $\pm$ 0.011 & 985.2 & 2464.8 & -0.88 $\pm$ 0.01 & 0.51 $\pm$ 0.02 \\
           &   &                              &                         & 343.2 & 829.5  & -0.24 $\pm$ 0.03 & 0.63 $\pm$ 0.05 \\
G34.26+0.15& M1 & 18 53 18.569 $\pm$ 0.007 & $+$01 14 57.997 $\pm$ 0.056 & 37.1  & 94.5  & 56.82 $\pm$ 0.06 & 0.68 $\pm$ 0.14 \\
           & M2 & 18 53 18.696 $\pm$ 0.002 & $+$01 14 55.807 $\pm$ 0.034 & 48.4  & 122.4 & 53.77 $\pm$ 0.05 & 0.35 $\pm$ 0.08 \\
           &   &                          &                             & 57.8  & 146.2 & 54.35 $\pm$ 0.07 & 0.83 $\pm$ 0.14 \\
           &   &                          &                             & 180.8 & 457.6 & 55.83 $\pm$ 0.01 & 0.59 $\pm$ 0.03 \\
           & M3 & 18 53 18.667 $\pm$ 0.005 & $+$01 14 55.348 $\pm$ 0.066 & 78.1  & 197.2 & 54.22 $\pm$ 0.04 & 0.94 $\pm$ 0.08 \\
           &   &                          &                             & 73.7  & 186.3 & 55.78 $\pm$ 0.04 & 0.79 $\pm$ 0.08 \\
\hline
\end{tabular}
%\tablefoot{}
\end{table}

\begin{figure*}[h]
\center
        \includegraphics[width=500pt]{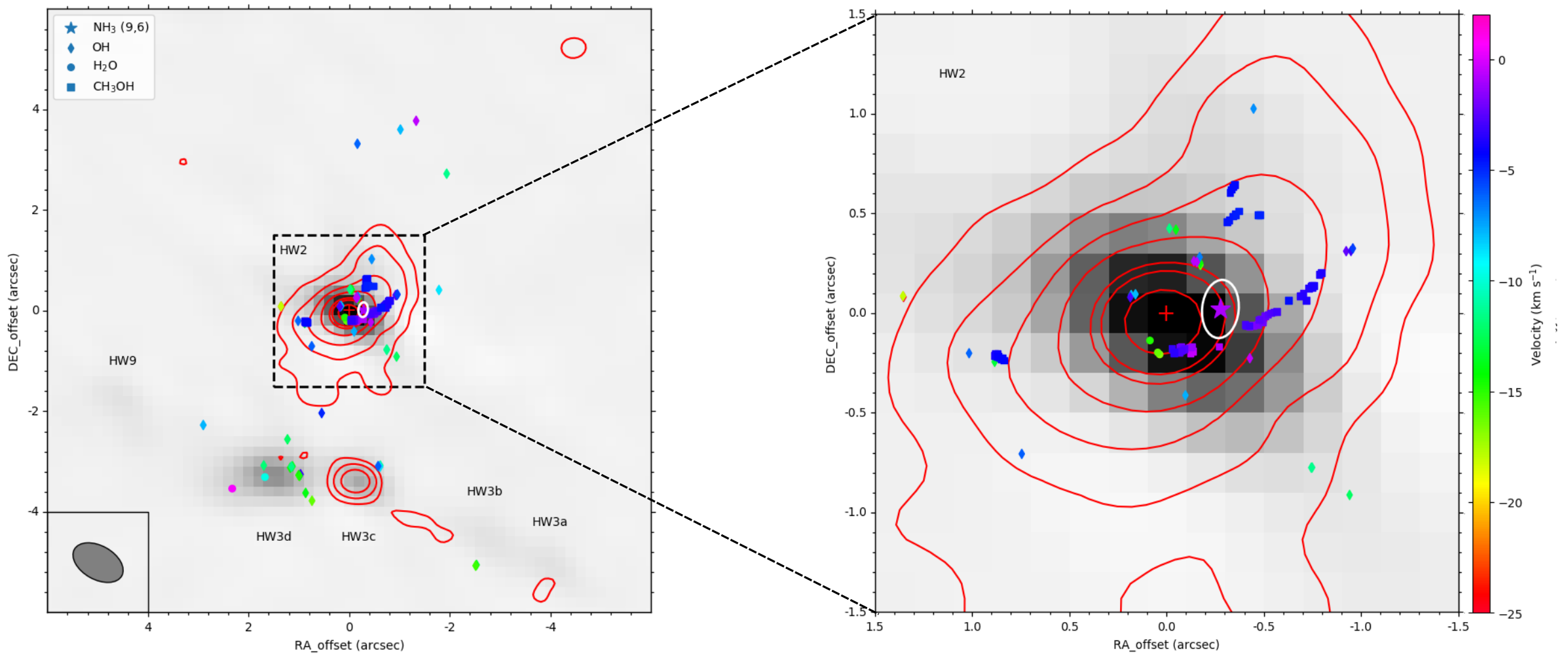}
  \caption{Cepheus A. The grey shaded areas mark the 1.36\,cm JVLA continuum map of Cep A. The reference position is $\alpha_{\rm J2000}$~=~22$^{\rm h}$56$^{\rm m}$17$\fs$972, and $\delta_{\rm J2000}$~=~62$\degr$01$\arcmin$49$\farcs$587, the peak position of the continuum map, is marked by a red cross. Slightly to the west of the cross is the white ellipse denoting the position of the NH$_3$ (9,6) emission with a purple star at its center. The red contours show the NOrthern Extended Millimeter Array (NOEMA) 1.37 mm continuum, taken from \citet{2018A&A...617A.100B}. Contour levels are -5, 5, 10, 20, 40, 80, 100, 150, and 200~$\times$~2.43~mJy~beam$^{-1}$. OH \citep{2005MNRAS.361..623B}, H$_2$O \citep{2018ApJ...856...60S}, and CH$_3$OH \citep{2017A&A...603A..94S} masers are presented as diamonds, circles, and squares, respectively. The color bar on the right-hand side indicates the velocity range ($V_{\rm LSR}$) of maser spots.}
  \label{cepa_continuum_cm_mm}
\end{figure*}

\begin{figure*}[h]
\center
        \includegraphics[width=500pt]{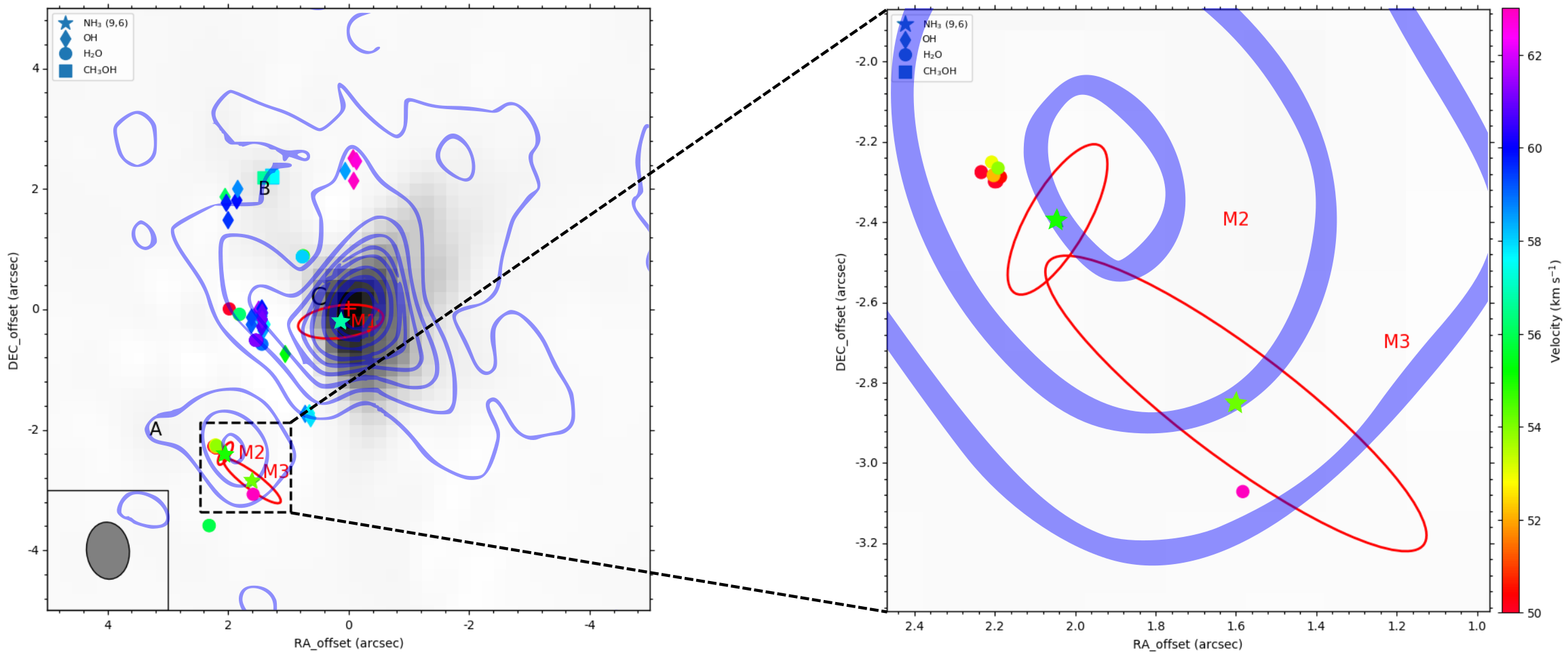}
  \caption{1.36\,cm JVLA continuum map of G34.26$+$0.15 presented as gray shaded areas. The reference position is $\alpha_{\rm J2000}$~=~18$^{\rm h}$53$^{\rm m}$18$\fs$560, and $\delta_{\rm J2000}$~=~01$\degr$14$\arcmin$58$\farcs$201, the peak position, is marked by a red cross. The red ellipses show the positions of NH$_3$ (9,6) emission with stars at their center (i.e., M1, M2, and M3). The blue contours show the Berkeley-Illinois-Maryland Association (BIMA) array 2.8 mm continuum, taken from \citet{2007ApJ...659..447M}. Contour levels are -3, 3, 10, 20, 30, 40, 50, 70, 90, 100, 120, and  140~$\times$~20~mJy~beam$^{-1}$. OH \citep{2000MNRAS.317..192Z}, H$_2$O \citep{2011PASJ...63.1293I}, and CH$_3$OH \citep{2016A&A...587A.104B} masers are presented as diamonds, circles, and squares, respectively. The color bar indicates the velocity range ($V_{\rm LSR}$) of maser spots.}
    \label{g34_continuum_cm_mm}
\end{figure*}
\end{appendix}

\end{document}